\documentstyle[12pt,subeqn,axodraw]{article}

\begin{document}
\title{Density Correlations in the Two-Dimensional Coulomb Gas}
\author{
L. {\v S}amaj$^{1,2}$ and B. Jancovici$^1$
}
\maketitle
\begin{abstract}
We consider a two-dimensional Coulomb gas of positive and negative
pointlike unit charges interacting via a logarithmic potential.
The density (rather than the charge) correlation functions
are studied.
In the bulk, the form-factor theory of an equivalent sine-Gordon model 
is used to determine the density correlation length.
At the surface of a rectilinear plain wall, the universality of 
the asymptotic behavior of the density correlations is suggested.
A scaling analysis implies a local form of the compressibility sum rule 
near a hard wall.
A symmetry of the Coulomb system with respect to the M\"obius conformal
transformation, which induces a gravitational source acting on the
particle density, is established.
Among the consequences, a universal term of the finite-size expansion 
of the grand potential is derived exactly for a disk geometry of
the confining domain.  
\end{abstract}

\smallskip


\smallskip

\noindent LPT Orsay 01-47

\vfill

\noindent $^1$ Laboratoire de Physique Th{\'e}orique, Universit{\'e} de
Paris-Sud, B{\^a}timent 210, 91405 Orsay Cedex, France (Unit{\'e} Mixte
de Recherche no. 8627 - CNRS); 

\noindent $^2$ On leave from the Institute of Physics, Slovak Academy of
Sciences, Bratislava, Slovakia

\noindent E-mail addresses: fyzimaes@savba.sk and 
Bernard.Jancovici@th.u-psud.fr
\newpage
 
\renewcommand{\theequation}{1.\arabic{equation}}
\setcounter{equation}{0}

\section{Introduction and summary}
In this paper, we consider classical Coulomb systems in 
thermodynamic equilibrium.
For the sake of simplicity, we will restrict ourselves to
the case of a symmetric two-component plasma (TCP), i.e.,
a neutral system of two species of particles of opposite
unit charges $\sigma=\pm 1$, living in a $\nu$-dimensional 
space and interacting through the Coulomb interaction
$\sigma_i \sigma_j v(\vert {\bf r}_i-{\bf r}_j \vert)$.
In dimension $\nu$, the Coulomb potential $v$ at a spatial
position ${\bf r} \in R^{\nu}$, induced by a unit charge
at the origin, is the solution of the Poisson equation
\begin{equation} \label{1.1}
\Delta v({\bf r}) = - s_{\nu} \delta({\bf r})
\end{equation}
where $s_{\nu}$ is the surface area of the
$\nu$-dimensional unit sphere; $s_2 = 2\pi$
and $s_3 = 4\pi$.
In particular,
\begin{subequations} \label{1.2}
\begin{eqnarray}
v({\bf r}) & = & - \ln \left( \vert {\bf r} \vert / r_0 \right) ,
\quad \quad \nu = 2 \label{1.2a} \\
v({\bf r}) & = & 1 / \vert {\bf r} \vert ,
\quad \quad \quad \quad \quad \ \nu = 3 \label{1.2b}
\end{eqnarray}
\end{subequations}
where, for $\nu = 2$, the length scale $r_0$ will be set
to unity without any loss of generality.
For the case of pointlike particles, the singularity of
$v({\bf r})$ at the origin prevents the thermodynamic
stability against the collapse of positive-negative pairs
of charges: in two dimensions for small enough temperatures,
in three dimensions for any temperature.

Let us now introduce some notations.
We will work in the grand canonical ensemble, characterized
by the fixed domain $D$ in which the TCP is confined, by the
inverse temperature $\beta$, and by the couple of equal
(constant) particle fugacities $z_+ = z_- = z$. 
The microscopic densities of charge and of the total
particle number are defined respectively by
\begin{equation} \label{1.3}
{\hat \rho}({\bf r}) = \sum_{\sigma} \sigma 
{\hat n}_{\sigma}({\bf r}) , \quad \quad
{\hat n}({\bf r}) = \sum_{\sigma}
{\hat n}_{\sigma}({\bf r})
\end{equation}
where ${\hat n}_{\sigma}({\bf r}) = \sum_i \delta_{\sigma,\sigma_i}
\delta({\bf r}-{\bf r}_i)$
is the microscopic density of particles of species 
$\sigma = \pm$ and $i$ indexes the particles.
The thermal average will be denoted by $\langle \cdots \rangle$.
At one particle level, the total charge and particle number
densities are given respectively by
\begin{equation} \label{1.4}
\rho({\bf r}) = \langle {\hat \rho}({\bf r}) \rangle ,
\quad \quad n({\bf r}) = \langle {\hat n}({\bf r}) \rangle
\end{equation}
Due to the charge $\pm$ symmetry, $n_+({\bf r}) =
n_-({\bf r}) = n({\bf r})/2$.
At two-particle level, one introduces the two-body densities
\begin{eqnarray} \label{1.5}
n_{\sigma\sigma'}({\bf r},{\bf r}') & = & 
\left\langle \sum_{i\ne j} \delta_{\sigma,\sigma_i}
\delta_{\sigma',\sigma_j} \delta({\bf r}-{\bf r}_i)
\delta({\bf r}'-{\bf r}_j) \right\rangle \nonumber \\
& = & \langle {\hat n}_{\sigma}({\bf r}) 
{\hat n}_{\sigma'}({\bf r}') \rangle - 
\langle {\hat n}_{\sigma}({\bf r}) \rangle 
\delta_{\sigma\sigma'} \delta({\bf r}-{\bf r}')
\end{eqnarray}
Clearly, $n_{++}({\bf r},{\bf r}') = n_{--}({\bf r},{\bf r}')$
and $n_{+-}({\bf r},{\bf r}') = n_{-+}({\bf r},{\bf r}')$.
The corresponding Ursell functions read
\begin{equation} \label{1.6}
U_{\sigma\sigma'}({\bf r},{\bf r}') = 
n_{\sigma\sigma'}({\bf r},{\bf r}') - 
n_{\sigma}({\bf r}) n_{\sigma'}({\bf r}')
\end{equation}
They will occur in the charge and density combinations
\begin{subequations} \label{1.7}
\begin{eqnarray}
U_{\rho}({\bf r},{\bf r}') & = & \sum_{\sigma,\sigma'}
\sigma \sigma' U_{\sigma\sigma'}({\bf r},{\bf r}') \label{1.7a} \\
U_{n}({\bf r},{\bf r}') & = & \sum_{\sigma,\sigma'}
U_{\sigma\sigma'}({\bf r},{\bf r}') \label{1.7b}
\end{eqnarray}
\end{subequations}
respectively.
The truncated charge-charge and density-density structure functions
are defined by
\begin{subequations} \label{1.8}
\begin{eqnarray} 
S_{\rho}({\bf r},{\bf r}') & = & \langle {\hat \rho}({\bf r})
{\hat \rho}({\bf r}') \rangle - 
\langle {\hat \rho}({\bf r}) \rangle 
\langle {\hat \rho}({\bf r}') \rangle \label{1.8a} \\
S_{n}({\bf r},{\bf r}') & = & \langle {\hat n}({\bf r})
{\hat n}({\bf r}') \rangle - 
\langle {\hat n}({\bf r}) \rangle 
\langle {\hat n}({\bf r}') \rangle \label{1.8b} 
\end{eqnarray}
\end{subequations}
respectively.
It is useful to consider also the pair correlation functions
\begin{equation} \label{1.9}
h_{\sigma\sigma'}({\bf r},{\bf r}') =
{n_{\sigma\sigma}({\bf r},{\bf r}') \over n_{\sigma}({\bf r})
n_{\sigma'}({\bf r}')} - 1
\end{equation}
in their charge and density combinations, defined by
\begin{subequations} \label{1.10}
\begin{eqnarray}
h_{\rho}({\bf r},{\bf r}') & = & {1\over 4} \sum_{\sigma,\sigma'}
\sigma \sigma' h_{\sigma\sigma'}({\bf r},{\bf r}') \label{1.10a} \\
h_{n}({\bf r},{\bf r}') & = & {1\over 4} \sum_{\sigma,\sigma'}
h_{\sigma\sigma'}({\bf r},{\bf r}') \label{1.10b}
\end{eqnarray}
\end{subequations}

As any Coulomb system, the TCP admits a description in 
the Debye-H\"uckel (high-temperature) limit \cite{Debye}.

In two dimensions and for the case of pointlike charged particles,
the particle density $n$ is an ``irrelevant'' variable which only
scales the distance.
The complete description of the bulk thermodynamics is available
in the whole range of inverse temperatures $\beta<2$ where the plasma 
is stable against the collapse of positive-negative pairs of charges.
The exact equation of state
\begin{equation} \label{1.11}
\beta p = n ( 1 - \beta/4)
\end{equation}
where $p$ is the pressure and $n$ the total particle density, 
has been known for a very long time \cite{Salzberg,Hauge}.
The evaluation of other thermodynamic quantities (free energy, 
internal energy, specific heat, etc.) can be based on
an explicit density-fugacity ($n-z$) relationship.
This relationship was obtained only recently \cite{Samaj1} via a 
mapping onto a classical two-dimensional sine-Gordon theory.
Surface thermodynamics (surface tension) of the two-dimensional
TCP in contact with a rectilinear dielectric wall was derived
for specific boundary conditions in refs. \cite{Samaj2, Samaj3}.
The large-distance behavior of the bulk charge-charge correlation
function was given in ref. \cite{Samaj4} by exploring the
form-factor method for the equivalent sine-Gordon theory.
Exact formulae for pair correlations for an arbitrary interparticle
distance are available just at the collapse point $\beta = 2$
\cite{Cornu1,Cornu2}, which corresponds to the free-fermion
point of an equivalent Thirring model.

The long-range tail of the Coulomb force causes screening,
and thus give rise to exact constraints for the charge correlations 
for any value of $\beta$ (for a review, see ref. \cite{Martin}).

In the bulk, the charge structure function
$S_{\rho}({\bf r},{\bf r}') = S_{\rho}(\vert {\bf r}
- {\bf r}' \vert)$ obeys the Stillinger-Lovett sum rules
\cite{Stillinger}, namely the zeroth-moment condition
\begin{equation} \label{1.12}
\int {\rm d}{\bf r} ~ S_{\rho}({\bf r}) = 0
\end{equation}
and the second-moment condition
\begin{equation} \label{1.13}
\int {\rm d}{\bf r} ~ \vert {\bf r} \vert^2 
S_{\rho}({\bf r}) =  - {\nu \over \beta \pi (\nu-1)},
\quad \quad \nu = 2, 3
\end{equation}
Hereinafter, without any loss of generality, we put
the dielectric constant $\epsilon$ of a medium in which
the plasma lives equal to unity.

Let us now introduce a semi-infinite TCP which occupies
the half space $x>0$, and denote by ${\bf y}$ the set of
$(\nu-1)$ coordinates normal to $x$.
The plane at $x=0$ is a hard wall impenetrable to the particles.
The half-space $x<0$ is assumed to be filled with a material
of dielectric constant $\epsilon_W$: a particle of unit charge
at the point ${\bf r} = (x>0,{\bf y})$ has an electric image
of charge $(1-\epsilon_W)/(1+\epsilon_W)$ at the
point ${\bf r}^* = (-x,{\bf y})$ \cite{Jackson}.
Due to invariance with respect to translations along the wall
and rotations around the $x$ direction,
\begin{equation} \label{1.14}
S_{\rho}({\bf r},{\bf r}') = 
S_{\rho}(x,x';\vert {\bf y}-{\bf y}' \vert) =
S_{\rho}(x',x;\vert {\bf y}-{\bf y}' \vert)
\end{equation}
The electroneutrality condition (\ref{1.12}) takes the form
\begin{equation} \label{1.15}
\int_0^{\infty} {\rm d}x' \int {\rm d}{\bf y} ~
S_{\rho}(x,x';{\bf y}) = 0
\end{equation}
The Carnie and Chan \cite{Carnie} generalization of the
second-moment condition (\ref{1.13}) results in the dipole
sum rule \cite{Blum,Jancovici0}
\begin{equation} \label{1.16}
\int_0^{\infty} {\rm d}x \int_0^{\infty} {\rm d}x'
\int {\rm d}{\bf y} ~ (x' - x) S_{\rho}(x,x';{\bf y})
= - {1 \over 2 \beta \pi (\nu - 1)},
\quad \quad \nu = 2, 3
\end{equation}
The asymmetry of the screening cloud of a charged particle
sitting near the wall induces a long-range tail in the
charge correlation along the wall \cite{Usenko,Jancovici1},
except of the cases $\epsilon_W=0$ (ideal dielectric wall) and
$\epsilon_W=\infty$ (ideal conductor wall) which will be
excluded from our considerations.
One expects an asymptotic power-law decay
\begin{equation} \label{1.17}
S_{\rho}(x,x';{\bf y}) \sim {f_{\rho}(x,x') \over
\vert {\bf y} \vert^{\nu}}, \quad \quad
\vert {\bf y} \vert \to \infty
\end{equation}
where $f_{\rho}(x,x')$ obeys the sum rule 
\cite{Jancovici2,Jancovici3}
\begin{equation} \label{1.18}
\int_0^{\infty} {\rm d}x \int_0^{\infty} {\rm d}x'
f_{\rho}(x,x') = - {\epsilon_W \over 2\beta 
[\pi (\nu-1)]^2}, \quad \quad \nu = 2, 3 
\end{equation}
Recently \cite{Jancovici4}, it was proven that for any value
of $x\ge 0$ it holds
\begin{equation} \label{1.19}
\int_0^{\infty} {\rm d}x' \int {\rm d}{\bf y} ~ (x' - x) 
S_{\rho}(x,x';{\bf y})
= {1 \over \epsilon_W} \pi (\nu-1) 
\int {\rm d}x' f_{\rho}(x,x'), \quad \quad \nu = 2, 3
\end{equation}
When both sides of (\ref{1.19}) are integrated over $x$ from 0
to $\infty$, it is clear that the sum rule (\ref{1.18}) for
$f_{\rho}$ is a direct consequence of the dipole sum 
rule (\ref{1.16}), and vice versa.

As concerns the density correlation function,
according to the general theory of fluids \cite{Hansen1},
the zeroth moment of the bulk density structure function
$S_n({\bf r},{\bf r}') = S_n(\vert {\bf r}-{\bf r}'\vert)$
is related to the isothermal compressibility
\begin{equation} \label{1.20}
\chi_{\beta} = {1\over n} \left( {\partial n \over
\partial p} \right)_{\beta}
\end{equation}
via
\begin{equation} \label{1.21}
{1\over n^2} \int {\rm d}{\bf r} ~ S_n({\bf r})
= {1\over \beta} \chi_{\beta}
\end{equation}

In two dimensions and for the case of pointlike charged particles,
with the use of the exact equation of state (\ref{1.11}), one
gets explicitly \cite{Vieillefosse}
\begin{equation} \label{1.22}
\int {\rm d}^2 r ~ S_n(r) = {n \over 1-(\beta/4)}
\end{equation}
Recently, using the technique of a renormalized Mayer
expansion \cite{Deutsch}, and in particular a remarkable
``cancelation property'' of specific families of renormalized
diagrams \cite{Kalinay}, the second moment of $S_n$ was shown
to have a simple value \cite{Jancovici5}
\begin{equation} \label{1.23}
\int {\rm d}^2 r ~ \vert {\bf r} \vert^2 S_n(r) = {1 \over 12 \pi 
\left( 1-(\beta/4) \right)^2}
\end{equation}

The universal finite-size properties of two-dimensional
critical systems with {\it short-range} interactions among
constituents are well understood within the principle of
conformal invariance \cite{Blote} - \cite{Cardy3}.
For a finite system of characteristic size $R$, at a critical
point, the dimensionless grand potential 
$\beta \Omega = - \ln \Xi$ ($\Xi$ is the grand partition function) 
has a large-$R$ expansion of the form
\begin{equation} \label{1.24}
\beta \Omega = A R^2 + B R - {c \chi \over 6} \ln R + 
{\rm const} + \cdots
\end{equation}
The coefficients $A$ and $B$ of the bulk and surface parts are
non-universal.
The coefficient of the $\ln R$-term is universal, dependent only
on the conformal anomaly number $c$ of the critical theory and
on the Euler number $\chi$ of the manifold on which the system
is confined.
In general, $\chi = 2 - 2h - b$, where $h$ is the number of handles
and $b$ the number of boundaries of the manifold ($\chi=2$ for
a sphere, $\chi=1$ for a disk, $\chi=0$ for an anulus or a torus).
The grand potential of the two-dimensional TCP is supposed to
exhibit a universal finite-size correction of type (\ref{1.24})
at any temperature of the conducting regime.
Plausible, but not always rigorously justified, arguments for a
critical-like behavior were first given for Coulomb gases
with periodic boundary conditions \cite{Forrester}, then
for Coulomb systems confined to a domain by plain hard walls
\cite{Jancovici6}, by ideal-conductor walls \cite{Jancovici7}
and finally by ideal-dielectric boundaries 
\cite{Jancovici8,Tellez}.
The explicit checks were done at the exactly solvable $\beta=2$
inverse temperature for various geometries of confining domains.
Only very recently \cite{Jancovici9}, a direct derivation of
the universal finite-size correction term was done for the
specific case of the TCP living on the surface of a sphere
of radius $R$.
By combining the method of stereographic projection of the sphere
onto an infinite plane with the linear response theory, 
the prefactor to the universal $\ln R$ correction term was related
to the bulk second moment of the density structure function
$S_n$, eq. (\ref{1.23}).
The obtained result confirms the prediction (\ref{1.24}) for
a Coulomb system, as if we had $c=-1$, in full agreement with 
heuristic approaches and exact results at $\beta=2$
\cite{Forrester}-\cite{Tellez}.

In general, due to screening phenomena, the sum rules
for the charge structure function $S_{\rho}$ are not modified by 
a short-distance regularization of the Coulomb potential.
On the other hand, the moments of the density structure 
function $S_n$ {\it depend} on the particular form of the
short-range particle interactions.
For the case of two dimensions and the pointlike character
of charged particles on which we shall concentrate in this
work, the zeroth-moment (\ref{1.12}) and the second-moment
(\ref{1.13}) Stillinger-Lovett conditions for $S_{\rho}$
have their exact counterparts (\ref{1.22}) and (\ref{1.23}),
respectively, for $S_n$.
Instead of screening, the scaling properties and the
critical-like state are relevant.
The aim of this paper is to document how these features
of the two-dimensional Coulomb fluids manifest themselves
in the density correlation functions (in the bulk or close to 
a boundary) and in the universal finite-size correction term 
of the grand potential.

Section 2 is devoted to the density correlations in the bulk.
The Debye-H\"uckel $\beta\to 0$ limit is presented in
subsection 2.1, the exactly solvable $\beta=2$ case is
treated in subsection 2.2.
A general analysis, based on the form-factor theory
of the equivalent sine-Gordon model in analogy with
ref. \cite{Samaj4}, is given in subsection 2.3.
For comparison, the corresponding formulae for the charge 
correlations are presented, too.

Density correlations near a rectilinear hard wall
are analysed in section 3.
In subsection 3.1., we present the results of the Debye-H\"uckel 
$\beta\to 0$ limit.
Subsection 3.2. deals with the exactly solvable $\beta=2$
case for the plain hard wall with $\epsilon_W=1$.
In comparison with formula (\ref{1.17}) for $S_{\rho}$ 
taken at dimension $\nu=2$, a more rapid, but still
power-law, asymptotic decay of $S_n$ along the boundary
is observed,
\begin{equation} \label{1.25}
S_n(x,x';y) \sim {f_n(x,x') \over y^4}
\end{equation} 
In subsection 3.3, at least for the considered plain hard wall, 
the universal form of the function
$f_n(x,x')$ at the boundary $x=x'=0$ is conjectured,
\begin{equation} \label{1.26}
f_n(0,0) = {1\over 2\pi^2}
\end{equation}
Here, we also suggest explicit forms of $f_{\rho}(x,x')$
and $f_n(x,x')$ for any value of $\beta$.

In section 4, we explore the scaling properties of the
two-dimensional TCP confined to a disk of radius $R$,
for the sake of simplicity by an uncharged plain hard wall.
We present some important formulae which are used in
the subsequent sections.
The rectilinear wall is obtained as the limiting $R\to\infty$
case of the disk.
As a by-product of the formalism, we derive a local form
of the compressibility (zeroth-moment) sum rule (\ref{1.22})
for $S_n$ near the rectilinear hard wall.

In section 5, by using a specific (namely M\"obius) conformal 
transformation of particle coordinates we show how the
two-dimensional TCP can be mapped onto the one under
the action of a gravitational source, acting in the
same way on both positively and negatively charged particles.
The mapping implies a new sum rule for $S_n$ in the disk geometry.
In this section, the new sum rule is used to derive the
counterpart of the bulk second-moment sum rule for $S_n$
(\ref{1.23}) near a rectilinear wall.

Section 6 deals with the universal finite-size correction
$\ln R$-term for the disk geometry.
Using the new sum rule for $S_n$ derived in section 5, 
the universal term is confirmed at any value of $\beta$.
This term has already been calculated at $\beta=2$ in
ref. \cite{Jancovici6}.

A brief recapitulation and some concluding remarks 
are given in section 7.

\renewcommand{\theequation}{2.\arabic{equation}}
\setcounter{equation}{0}

\section{Bulk density correlations}

\subsection{Weak coupling}
In this subsection, we derive the asymptotic form of the density
correlation function $h_n(r)$ in the bulk, at the lowest order in
$\beta$. 
It should be noted that $\beta$-expansions of correlation
functions must be taken for a fixed value of the inverse Debye length
$\kappa=(2\pi\beta n)^{1/2}$, since $\kappa$ only fixes the length
scale.

In the present case of a charge-symmetrical two-component
plasma, the ordinary Ornstein-Zernike (OZ) equation split
into two independent relations for the charge-charge and
density-density functions \cite{Jancovici5}
\begin{subequations} \label{2.1}
\begin{eqnarray} 
h_{\rho} & = & c_{\rho} + c_{\rho} \ast n \ast h_{\rho}
\label{2.1a} \\
h_n & = & c_n + c_n \ast n \ast h_n \label{2.1b}
\end{eqnarray}
\end{subequations}
respectively, where $\ast$ denotes a convolution product.
In the renormalized Mayer expansion \cite{Samaj1,Jancovici5}, 
at the lowest order in $\beta$, the charge direct correlation
function is given by
\begin{equation} \label{2.2}
c_{\rho}(1,2) = \ \ \
{\begin{picture}(40,20)(0,7)
    \DashLine(0,10)(40,10){5}
    \BCirc(0,10){2.5} \BCirc(40,10){2.5}
    \Text(0,0)[]{1} \Text(40,0)[]{2}
\end{picture}}\ \ = - \beta v(1,2)
\end{equation}
with $v$ being the Coulomb potential, and the density direct 
correlation function is given by the Meeron diagram
\begin{equation} \label{2.3}
c_n(1,2) =  \ \ \
\begin{picture}(40,25)(0,7)
    \PhotonArc(20,-10)(28,45,135){1}{9}
    \PhotonArc(20,30)(28,225,315){1}{9}
    \BCirc(0,10){2.5} \BCirc(40,10){2.5}
    \Text(0,0)[]{1} \Text(40,0)[]{2}
\end{picture}\ \ = {1\over 2!} K^2(1,2)
\end{equation}
The wavy line denotes the renormalized bond $K$ (sum of chains).
In the bulk, $K(r) = - \beta K_0(\kappa r)$, where $K_0$ 
is a modified Bessel function.

Inserting (\ref{2.2}) into (\ref{2.1a}), one gets
the definition of $K$.
This is why
\begin{equation} \label{2.4}
h_{\rho}(r) = - \beta K_0(\kappa r) \sim
- \beta \left( {\pi \over 2\kappa r} \right)^{1/2}
\exp ( - \kappa r )
\end{equation}
since $K_0(x)$ has the asymptotic form $[\pi/(2x)]^{1/2}\exp(-x)$. 
In the OZ relation (\ref{2.1b}), where $c_n$ (\ref{2.3}) 
is of order $\beta^2$, the convolution term is easily seen 
to be of higher order $\beta^3$. 
Thus, at lowest order in $\beta$, $h_n(r) = c_n(r)$, 
\begin{equation} \label{2.5}
h_n(r) = {\beta^2 \over 2} K_0^2(\kappa r) \sim
{\pi\beta^2 \over 4 \kappa r} \exp (-2 \kappa r)
\end{equation}

The same analysis can be done in dimension $\nu = 3$,
where
\begin{equation} \label{2.6}
K(r) = - { \beta \over r} \exp( -\kappa r) 
\end{equation}
where now $\kappa = (4\pi \beta n)^{1/2}$ \cite{Lee}.
In this case, the large-$r$ behavior is not touched by an
inevitable short-distance regularization of the Coulomb potential.

\subsection{$\beta$=2}
When $\beta=2$, for a fixed fugacity $z$, the particle density
$n\to \infty$ and $\{ h_{\rho}, h_n \} \to 0$.
However, the Ursell functions $U_{\rho} = n^2 h_{\rho}$ and
$U_n = n^2 h_n$ are exactly known \cite{Cornu1,Cornu2} as
\begin{equation} \label{2.7}
U_{\rho}(r) = - 2 \left( \frac{m^2}{2\pi} \right)^2
\left[ K_1^2(mr) + K_0^2(mr) \right]
\end{equation}
and
\begin{equation} \label{2.8}
U_n(r) = 2 \left( \frac{m^2}{2\pi} \right)^2
\left[ K_1^2(mr) - K_0^2(mr) \right]
\end{equation}
where $m=2\pi z$ ($z$ is the fugacity). 
Replacing the modified Bessel functions
by their asymptotic expansions
\begin{subequations} \label{2.9}
\begin{eqnarray}
& & K_0(x)=\left(\frac{\pi}{2x}\right)^{1/2}
{\rm e}^{-x}\left[1-\frac{1}{8x}+\ldots \right] \label{2.9a} \\
& & K_1(x)=\left(\frac{\pi}{2x}\right)^{1/2}
{\rm e}^{-x}\left[1+\frac{3}{8x}+\ldots \right] \label{2.9b}
\end{eqnarray}
\end{subequations}
gives the asymtotic form of $U_{\rho}(r)$ as
\begin{equation} \label{2.10}
U_{\rho}(r) \sim - {m^3\over 2\pi r} \exp ( -2 m r)
\end{equation} 
and the asymtotic form of $U_n(r)$ as
\begin{equation} \label{2.11}
U_n(r) \sim  {m^2\over 4\pi r^2} \exp ( -2 m r)
\end{equation} 

\subsection{Form-factor analysis}
We now determine the large-distance behavior of $h_n(r)$ 
in the whole stability interval of inverse
temperatures $0<\beta<2$ via a form-factor analysis.

The grand partition function of the TCP can be turned into
(see, e.g., \cite{Minnhagen})
\begin{equation} \label{2.12}
\Xi = { \int {\cal D}\phi ~ \exp \left( - S(z) \right)
\over \int {\cal D}\phi ~ \exp \left( - S(0) \right)}
\end{equation}
where 
\begin{subequations} \label{2.13}
\begin{eqnarray}
S(z) & = & \int {\rm d}^2 r \left[ {1\over 16\pi} \left(
\nabla \phi \right)^2 - 2 z \cos (b\phi) \right] \label{2.13a} \\
b^2 & = & \beta / 4 \label{2.13b}
\end{eqnarray}
\end{subequations}
is the Euclidean action of the classical sine-Gordon model.
In the sine-Gordon representation, the density of particles
of one sign $\sigma = \pm$ is
\begin{equation} \label{2.14}
n_{\sigma} = z_{\sigma} \langle 
{\rm e}^{{\rm i}\sigma b \phi} \rangle
\end{equation}
where $\langle \cdots \rangle$ denotes the averaging over the
sine-Gordon action (\ref{2.13}),
two-body densities (\ref{1.5}) are expressible as follows
\begin{equation} \label{2.15}
n_{\sigma,\sigma'}({\bf r},{\bf r}') = z_{\sigma} z_{\sigma'}
\langle {\rm e}^{{\rm i}\sigma b \phi({\bf r})}
{\rm e}^{{\rm i}\sigma' b \phi({\bf r}'} \rangle
\end{equation}
etc.
The parameter $z=z_+=z_-$, which is the fugacity renormalized by
a (diverging) self-energy term, gets a precise meaning when one
fixes the normalization of the cos-field.
For the TCP \cite{Samaj1}, this normalization is given by the
short-distance behavior
\begin{equation} \label{2.16}
n_{+-}({\bf r},{\bf r}') \sim z_+ z_- \vert {\bf r}-{\bf r}'
\vert^{-\beta} \quad \quad 
{\rm as}\ \vert {\bf r}-{\bf r}'\vert \to 0
\end{equation}
dominated by the Boltzmann factor of the Coulomb potential.
The corresponding formula in the sine-Gordon picture
\begin{equation} \label{2.17}
\langle {\rm e}^{{\rm i}b \phi({\bf r})}
{\rm e}^{-{\rm i}b \phi({\bf r}'} \rangle \sim 
\vert {\bf r}-{\bf r}' \vert^{-4 b^2} \quad \quad 
{\rm as}\ \vert {\bf r}-{\bf r}'\vert \to 0
\end{equation}
is known in quantum field theory as the conformal normalization.

The sine-Gordon model (\ref{2.13}) is massive in the region
$0 < b^2 < 1$ $(0 < \beta < 4)$.
It is integrable \cite{Zamolodchikov1}:
its particle spectrum consists of one soliton-antisoliton pair
of equal masses $M$ and of soliton-antisoliton bound states
(``breathers'') $\{ B_j; j = 1, 2,\ldots < 1/\xi \}$.
Their number depends on the inverse of the parameter
\begin{equation} \label{2.18}
\xi = {b^2\over 1-b^2} \quad \quad 
\left( = {\beta \over 4-\beta} \right)
\end{equation}
The mass of the $B_j$-breather is given by
\begin{equation} \label{2.19}
m_j = 2 M \sin \left( {\pi \xi \over 2} j \right)
\end{equation}
and the breather disappears from the spectrum just when
$m_j = 2 M$.
Under the conformal normalization (\ref{2.17}), the relationship
between the soliton mass $M$ and the parameter $z$ was established
in \cite{Zamolodchikov2},
\begin{equation} \label{2.20}
z = {\Gamma(b^2) \over \pi \Gamma(1-b^2)} 
\left[ M {\sqrt{\pi} \Gamma((1+\xi)/2) \over 2 \Gamma(\xi/2)}
\right]^{2-2b^2}
\end{equation}
where $\Gamma$ stands for the Gamma function.
Using the Thermodynamic Bethe ansatz, the specific quantity
\begin{equation} \label{2.21}
\lim_{V\to\infty} {1\over V} \ln \Xi = { m_1^2 \over
8 \sin(\pi \xi)}
\end{equation}
was found in ref. \cite{Destri}.
As a thermodynamic result \cite{Samaj1},
\begin{equation} \label{2.22}
n = {M^2 \over 4 (1-b^2)} {\rm tan}\left( {\pi \xi \over 2} \right)
\end{equation}
Note that as $\beta$ approaches the collapse value 2, for a fixed $z$, 
$M$ is finite and $n\to\infty$ as it should be.

For the underlying sine-Gordon theory, the two-point truncated 
correlation functions of local operators ${\cal O}_a$ 
($a$ is a free parameter) can be formally written as an infinite 
convergent series over multi-particle intermediate states,
\begin{eqnarray}
\langle {\cal O}_a({\bf r}) {\cal O}_{a'}({\bf r}') \rangle_{{\rm T}}
& = & \sum_{N=1}^{\infty} {1\over N!} \sum_{\epsilon_1,\ldots,
\epsilon_N} \int_{-\infty}^{\infty} {{\rm d}\theta_1 \ldots
{\rm d}\theta_N \over (2\pi)^N}
F_a(\theta_1,\ldots,\theta_N)_{\epsilon_1\ldots\epsilon_N}
\nonumber \\ & & \times ~
^{\epsilon_N\ldots\epsilon_1}F_{a'}(\theta_N,\ldots,\theta_1)
\exp \left( - \vert {\bf r} - {\bf r}' \vert \sum_{j=1}^N
m_{\epsilon_j} \cosh \theta_j \right) \label{2.23}
\end{eqnarray}
where $\epsilon$ indexes the particles [$\epsilon = + (-)$
for a soliton (antisoliton) and $\epsilon = j$ for a breather
$B_j$ $(j=1,2,\ldots)$] and $\theta$ is the particle rapidity.
The form factors
\begin{subequations} \label{2.24}
\begin{eqnarray}
F_a(\theta_1,\ldots,\theta_N)_{\epsilon_1\ldots\epsilon_N}
& = & \langle 0 \vert {\cal O}_a({\bf 0}) \vert 
Z_{\epsilon_1}(\theta_1),\ldots,Z_{\epsilon_N}(\theta_N) \rangle
\label{2.24a} \\
^{\epsilon_N\ldots \epsilon_1}F_{a'}(\theta_N,\ldots,\theta_1)
& = & \langle Z_{\epsilon_N}(\theta_N),\ldots,Z_{\epsilon_1}(\theta_1) 
\vert {\cal O}_{a'}({\bf 0}) \vert 0 \rangle \label{2.24b}
\end{eqnarray}
\end{subequations}
are the matrix elements of the operator at the origin, between
an $N$-particle in-state (as a linear superposition of free
one-particle states $\vert Z_{\epsilon}(\theta)\rangle$)
and the vacuum.

In the limit $\vert {\bf r}-{\bf r}' \vert \to \infty$,
the dominant contribution to the truncated correlation function
in (\ref{2.23}) comes from a multi-particle state with the
minimum value of the total particle mass $\sum_{j=1}^N m_{\epsilon_j}$,
at the point of vanishing rapidities $\theta_j \to 0$.
Due to topological reasons, solitons and antisolitons coexist
in pairs, the total mass of the pair being $2 M$.
The breathers $B_j$ with masses given by relation (\ref{2.19})
are lighter and therefore, when they exist and their form-factor
contributions do not vanish, they are the best candidates for
governing the asymptotic behavior of the two-point correlation
function.

The form factors of an exponential operator
${\cal O}_a = \exp({\rm i}a \phi({\bf r}))$
for various combinations of particles
were calculated in refs. \cite{Smirnov} - \cite{Lukyanov2}.
The one-breather form factors, which do not depend
on the rapidity, posses the following general structure
\begin{equation} \label{2.25}
\langle 0 \vert {\rm e}^{{\rm i}\sigma b\phi} \vert
B_j(\theta) \rangle = 
\langle B_j(\theta) \vert {\rm e}^{{\rm i}\sigma b\phi} \vert
0 \rangle \propto \sigma^j 
\langle {\rm e}^{{\rm i}\sigma b\phi} \rangle 
\sin \left( {\pi j \over 2} \right)
\end{equation}
where the full dependence on $\sigma = \pm 1$ is presented.
These form factors are nonzero only if $j$ $=$ odd integer.
Due to the invariance of the sine-Gordon action (\ref{2.13})
with respect to the transformation, it holds
$\langle {\rm e}^{{\rm i}b\phi} \rangle =
\langle {\rm e}^{-{\rm i}b\phi} \rangle$.
With regard to (\ref{2.15}), the total contribution of
a given breather $B_j$ ($j$ odd) to the charge $h_{\rho}$
and density $h_n$ correlation functions (\ref{1.10})
is proportional to
\begin{subequations} \label{2.26}
\begin{eqnarray}
h_{\rho} & \propto & \sum_{\sigma,\sigma'=\pm 1} \sigma \sigma' 
\langle 0 \vert {\rm e}^{{\rm i}\sigma b\phi} \vert B_j(\theta) \rangle 
\langle B_j(\theta) \vert {\rm e}^{{\rm i}\sigma' b\phi}\vert 0 \rangle
\label{2.26a} \\ 
h_n & \propto & \sum_{\sigma,\sigma'=\pm 1}
\langle 0 \vert {\rm e}^{{\rm i}\sigma b\phi} \vert B_j(\theta) \rangle 
\langle B_j(\theta) \vert {\rm e}^{{\rm i}\sigma' b\phi}\vert 0 \rangle
\label{2.26b}
\end{eqnarray}
\end{subequations}

Inserting (\ref{2.25}) into (\ref{2.26}) one observes that
one-breather states contribute only to $h_{\rho}$.
Using (\ref{2.22}) in (\ref{2.19}), the mass of the lightest
(elementary) $B_1$-breather is
\begin{equation} \label{2.27}
m_1 = \kappa \left[ {\sin (\pi \beta/(4-\beta)) \over
\pi \beta/(4-\beta)} \right]^{1/2}
\end{equation}
where $\kappa$ denotes as usually the inverse Debye length.
At asymptotically large $r$, (\ref{2.23}) gives
\begin{equation} \label{2.28}
h_{\rho}(r) \propto \exp (-m_1 r) , \quad \quad
0 < \beta < 2
\end{equation}
in agreement with the $\beta\to 0$ limit (\ref{2.4}).
At the free-fermion point $\beta=2$, the $B_1$-breather disappears,
and the soliton-antisoliton pair with mass $2 M$ determines the
correlation length:
\begin{equation} \label{2.29}
U_{\rho}(r) \propto \exp (- 2 M r) , \quad \quad
\beta = 2
\end{equation}
From (\ref{2.20}), $M = 2\pi z = m$ at $\beta=2$ $(b^2=1/2)$, and the
asymptotic form (\ref{2.10}) is reproduced.
Note that from (\ref{2.19}) $m_1 \to 2M$ as $\beta\to 2$,
and so the inverse correlation length varies continuously near $\beta=2$.  
The explicit inverse-power law dependence of prefactors of asymptotic 
formulae (\ref{2.28}) 
and (\ref{2.29}) was presented in ref. \cite{Samaj4}.

At small $\beta$, the large-distance behavior of $h_n$ is
determined by the two-$B_1$-breather state.
Indeed, the corresponding form factor \cite{Samaj4,Lukyanov1}
\begin{equation} \label{2.30}
\langle 0 \vert {\rm e}^{{\rm i}\sigma b\phi} 
\vert B_1(\theta_2),B_1(\theta_1) \rangle = 
\langle B_1(\theta_2),B_1(\theta_1) \vert {\rm e}^{{\rm i}\sigma b\phi} 
\vert 0 \rangle \propto \sigma^2 
\langle {\rm e}^{{\rm i}\sigma b\phi} \rangle 
\end{equation}
(where the full dependence on $\sigma = \pm 1$ is given)
has the necessary $\sigma\to -\sigma$ symmetry and thus
contributes to $h_n$.
It follows from (\ref{2.19}) that the mass of two $B_1$-breathers, 
$2 m_1$, is smaller than the one of the soliton-antisoliton pair, $2 M$,
in the region $\beta<1$.
Consequently,
\begin{equation} \label{2.31}
h_n(r) \propto \exp ( -2 m_1 r) , \quad \quad 0 < \beta < 1
\end{equation}
at large distance $r$, in agreement with 
the $\beta\to 0$ limit (\ref{2.5}).
It stands to reason that subsequently the soliton-antisoliton
pair determines the large-distance asymptotic of $h_n$,
\begin{equation} \label{2.32}
h_n(r) \propto \exp ( -2 M r) , \quad \quad 1 \le \beta \le 2
\end{equation}
where the $\beta$-dependence of $M$ can be deduced from 
eq. (\ref{2.22}).
At $\beta=2$, the result (\ref{2.11}) is recovered.
We omit a tedious calculation of the inverse-power law dependence
of prefactors on distance in (\ref{2.31}) and (\ref{2.32}).

One concludes that, for a given $\beta < 2$, the large-distance
exponential decay of $h_n$ is faster than the one of $h_{\rho}$.
Although the correlation lengths depend continuously on $\beta$ 
for both $h_{\rho}$ and $h_n$, the derivative of the density correlation 
length with respect to $\beta$ is discontinuous at $\beta=1$.
The correlation lengths coincide just at the collapse $\beta=2$ point, 
where, as is clear from the exact formulae (\ref{2.10}) and
(\ref{2.11}), $h_{\rho}$ and $h_n$ differ from one another only by
inverse-power law prefactors. 

\renewcommand{\theequation}{3.\arabic{equation}}
\setcounter{equation}{0}

\section{Density correlations near a rectilinear hard wall}

\subsection{Weak coupling}
Like in Section 2.1, at the lowest order in $\beta$, 
the direct correlation functions $c_{\rho}$ and $c_n$
are given by relations (\ref{2.2}) and (\ref{2.3}), respectively.
However, now, the wavy line $K$, the sum of the chain diagrams, 
is a function $K(x,x';|y-y'|)$ which must be calculated with 
the wall taken into account. 
An arbitrary $\epsilon_W$ value can be assumed for 
the dielectric constant of the wall material. 
At the lowest order in $\beta$, $K$ can be computed with the
density $n(x)$ replaced by its bulk value $n$. 
This $K$ has been implicitly considered in refs. 
\cite{Jancovici4,Jancovici10}, where it was used as the charge 
correlation function, $h_{\rho} = K$. 
Except in the cases $\epsilon_W=\infty$ (ideal conductor wall) 
and $\epsilon_W=0$ (ideal dielectric wall), $K$ has an algebraic decay, 
for large $|y-y'|$, of the form 
$-\epsilon_W/[\pi n(y-y')^2]\exp[-\kappa (x+x')]$. 
Therefore, 
\begin{equation} \label{3.1}
U_{\rho}(x,x';|y-y'|) = n^2 K(x,x';|y-y'|) 
\sim - { \epsilon_W n \over \pi (y-y')^2 }
\exp \left[ - \kappa (x+x') \right] 
\end{equation}
The Meeron diagram (\ref{2.3}) gives for the asymptotic form 
of the density Ursell function 
\begin{equation} \label{3.2}
U_n(x,x';|y-y'|) = n^2 c_n(x,x';|y-y'|) 
\sim {\epsilon_W^2 \over 2\pi^2(y-y')^4}
\exp \left[ -2 \kappa (x+x') \right] 
\end{equation}
These formulae can be generalized to $\nu = 2, 3$ dimensions:
\begin{eqnarray}
U_{\rho}(x,x';|{\bf y}|) & \sim & 
- { \epsilon_W n \over (\nu -1) \pi \vert {\bf y} \vert^{\nu} }
\exp \left[ - \kappa (x+x') \right] \label{3.3} \\
U_n(x,x';|{\bf y}|) & \sim & 
{\epsilon_W^2 \over 2 (\nu -1)^2 \pi^2 \vert {\bf y} \vert^{2\nu}}
\exp \left[ -2 \kappa (x+x') \right] \label{3.4}
\end{eqnarray}
with $\kappa = [ 2 (\nu-1) \pi \beta n]^{1/2}$.

\subsection{$\beta$=2}
When $\beta$=2 and $\epsilon_W=1$, the charge and density Ursell 
functions can be expressed in terms of auxiliary functions 
$g_{\sigma +}$, where $\sigma = \pm $, as \cite{Cornu2}
\begin{subequations} \label{3.5}
\begin{eqnarray}
U_{\rho}(x,x';|y-y'|) & = & 
- 2 m^2 \left[ |g_{-+}(x,x';y-y')|^2 + |g_{++}(x,x';y-y')|^2 \right]
\label{3.5a} \\
U_n(x,x';|y-y'|) & = & 
2 m^2 \left[ |g_{-+}(x,x';y-y')|^2 - |g_{++}(x,x';y-y')|^2 \right]
\label{3.5b}
\end{eqnarray}
\end{subequations}
where $m=2\pi z$ ($z$ is the fugacity). 
It will now be shown that, near the wall, each $g$ function has a slow 
algebraic decay for large $|y-y'|$. 

Each $g$ is a sum of two terms
\begin{equation} \label{3.6}
g_{\sigma +}(x,x';y-y') = g_{\sigma +}^{\rm bulk}({\bf r}-{\bf r}')
+ g_{\sigma +}^{\rm wall}(x,x';y-y')
\end{equation}
The first term in the rhs of (\ref{3.6}) is the same as in the bulk. 
It has a fast (exponential) decay and does not contribute to 
the asymptotic form, which is entirely due to the second term. 
This second term is determined by its Fourier transform as
\begin{equation} \label{3.7}
g_{\sigma +}^{\rm wall}(x,x';y-y') =
\int_{-\infty}^{\infty} \frac{{\rm d}l}{2\pi} ~
\tilde{g}_{\sigma +}^{\rm wall}(x,x';l)\exp[{\rm i}l(y-y')]
\end{equation}
These Fourier transforms are
\begin{subequations} \label{3.8}
\begin{eqnarray}
\tilde{g}_{++}^{\rm wall}(x,x';l)&=&
-\frac{m}{2k}\exp[-k(x+x')],\; l<0 \label{3.8a} \\
\tilde{g}_{++}^{\rm wall}(x,x';l)&=&
\frac{m(k-l)}{2k(k+l)}\exp[-k(x+x')],\; l>0 \label{3.8b} \\
\tilde{g}_{-+}^{\rm wall}(x,x';l)&=&
-\frac{k+l}{2k}\exp[-k(x+x')],\; l<0 \label{3.8c} \\
\tilde{g}_{-+}^{\rm wall}(x,x';l)&=&
\frac{k-l}{2k}\exp[-k(x+x')],\; l>0 \label{3.8d}
\end{eqnarray}
\end{subequations}
where $k=(m^2+l^2)^{1/2}$. 
Thus, $\tilde{g}_{\sigma +}^{\rm wall}(x,x';l)$ 
is a function of $l$ singular at $l=0$. 
This singularity generates in $g_{\sigma +}^{\rm wall}(x,x';y-y')$ 
an algebraic decay at large $|y-y'|$. 
The corresponding asymptotic expansion is obtained by splitting 
the integral in (\ref{3.7}) into the $l<0$ and $l>0$ contributions, 
and evaluating each contribution by successive integration per partes. 
For instance, calling for simplicity $N(l)$ the function
$\tilde{g}_{\sigma +}^{\rm wall}(x,x';l)$ when $l<0$, one obtains the 
asymptotic expansion
\begin{eqnarray} \label{3.9}
& & \int_{-\infty}^0\frac{{\rm d}l}{2\pi}N(l)
\exp[{\rm i}l(y-y')] \nonumber \\
& & =\frac{1}{2\pi}\left\{-\frac{{\rm i}}{y-y'}N(0)+
\frac{1}{(y-y')^2}N'(0)+\frac{{\rm i}}{(y-y')^3}N''(0)+\ldots \right\}
\end{eqnarray}
A similar expansion holds for $P(l)$, the function 
$\tilde{g}_{\sigma +}^{\rm wall}(x,x';l)$ when $l>0$. 
Using these expansions finally gives
\begin{subequations} \label{3.10}
\begin{eqnarray} \label{3.10a}
|g_{++}(x,x';y-y')|^2 & = & \frac{1}{4\pi^2}\left\{\frac{1}{(y-y')^2}
+ \left[ -\frac{1}{m^2}+\frac{2(x+x')}{m} \right]
\frac{1}{(y-y')^4}+\ldots\right\} \nonumber \\
& & \times \exp[-2m(x+x')]
\end{eqnarray}
and
\begin{equation} \label{3.10b}
|g_{-+}(x,x';y-y')|^2 = \frac{1}{4\pi^2}\left\{\frac{1}{(y-y')^2}
+\frac{2(x+x')}{m(y-y')^4} + \ldots\right\} \exp[-2m(x+x')] 
\end{equation}
\end{subequations}
 
The dominant term in eqs. (\ref{3.10}), of order $1/(y-y')^2$, 
determines the asymptotic form of the charge Ursell function 
(\ref{3.5a}) \cite{Jancovici4},
\begin{equation} \label{3.11}
U_{\rho}(x,x';|y-y'|) \sim - \left( {m\over \pi} \right)^2
{1\over (y-y')^2} \exp \left[ -2m(x+x') \right]
\end{equation}
But this dominant term cancels out in the expression (\ref{3.5b}) 
of the density Ursell function, the asymptotic form of which 
is governed by the subdominant term in eqs. (\ref{3.10}), 
of order $1/(y-y')^4$. 
The final result is
\begin{equation} \label{3.12}
U_n(x,x';|y-y'|) \sim \frac{1}{2\pi^2(y-y')^4}
\exp[-2m(x+x')]
\end{equation}

\subsection{Conjectures}
The functions $U_{\rho,n}(r)$ and $S_{\rho,n}(r)$ differ from
one another only by a term containing $\delta(r)$, which
has no effect on their identical large-distance behavior.
With regard to the definitions of the asymptotic characteristics
$f_{\rho}$ (\ref{1.17}) and $f_n$ (\ref{1.25}), the results
of the two previous subsections can be summarized, for $\nu=2$
dimensions and $\epsilon_W=1$, by
\begin{subequations} \label{3.13}
\begin{eqnarray}
f_{\rho}(x,x') & = & - {n\over \pi} \exp \left[
- \kappa (x+x') \right] , \quad \quad \beta \to 0
\label{3.13a} \\
f_{\rho}(x,x') & = & -  \left( {m\over \pi} \right)^2
\exp \left[ - 2 m (x+x') \right] , \quad \quad \beta = 2
\label{3.13b}
\end{eqnarray}
\end{subequations} 
and
\begin{subequations} \label{3.14}
\begin{eqnarray}
f_n(x,x') & = & {1\over 2 \pi^2} \exp \left[
- 2 \kappa (x+x') \right] , \quad \quad \beta \to 0
\label{3.14a} \\
f_n(x,x') & = & {1\over 2 \pi^2} \exp \left[ 
- 2 m (x+x') \right] , \quad \quad \beta = 2
\label{3.14b}
\end{eqnarray}
\end{subequations} 

The explicit forms of $f_{\rho}$ and $f_n$ in the two
exactly solvable cases has an appealing feature:
both charge and density functions factorize in
$x$ and $x'$ particle coordinates.
This can be intuitively explained by the fact
that one studies the leading asymptotic $y\to\infty$
limit of pair correlations along the wall, in which
probably there is no correlation between $x$ and $x'$
coordinates of particles.
In both $\beta\to 0$ and $\beta=2$ cases, the decay
of factorized functions into the bulk has the exponential
form obtained for the corresponding bulk functions in
subsections 2.1. and 2.2.
With regard to the general form-factor analysis in
subsection 2.3., it is therefore tempting to write down
\begin{equation} \label{3.15}
f_{\rho}(x,x') = - {m_1^2 \over 2 \beta \pi^2}
\exp \left[ - m_1 (x+x') \right] , \quad \quad
0 < \beta \le 2
\end{equation}
with the mass $m_1$ of the elementary $B_1$-breather 
given by formula (\ref{2.27}).
This formula reproduces correctly the two solvable cases
(\ref{3.13}) and reflects the dominance of the lightest
particle in the sine-Gordon spectrum in the large-distance
behavior of charge correlation functions.
The value of the prefactor in (\ref{3.15}) is fixed by
the sum rule (\ref{1.18}) with $\nu=2$ and $\epsilon_W=1$.
As concerns the density asymptotic charasteristic $f_n(x,x')$
(\ref{3.14}), the prefactor of the exponential acquires
the same value in the $\beta\to 0$ limit as well as at $\beta=2$,
what motivates us to suggest, taking into account large-distance
asymptotics (\ref{2.31}) and (\ref{2.32}), that
\begin{subequations} \label{3.16}
\begin{eqnarray}
f_n(x,x') & = & {1\over 2\pi^2} \exp \left[
- 2 m_1 (x+x') \right] , \quad \quad 0 < \beta < 1
\label{3.16a} \\
f_n(x,x') & = & {1\over 2\pi^2} \exp \left[
- 2 M (x+x') \right] , \quad \quad 1 < \beta \le 2
\label{3.16b}
\end{eqnarray}
\end{subequations}  
In particular, at the boundary $x = x' = 0$,
$f_n(0,0)$ is supposed to have a universal value $1/(2\pi^2)$
independent of $\beta$.
We were not able to give some general argument for such a result.

We emphasize that formulae (\ref{3.15}) and (\ref{3.16}) are
only conjectures which must be verified, for example, 
within a systematic weak-coupling expansion in the presence 
of a dielectric wall.
Such calculations for the correlation functions are tedious 
and far from simple, and will not be presented here.

\renewcommand{\theequation}{4.\arabic{equation}}
\setcounter{equation}{0}

\section{Scaling analysis in a disk}
Under the neutrality constraint $N_+ = N_-$, the grand partition
function of the two-dimensional symmetric TCP in a domain $D$,
bounded by an impermeable hard wall (for simplicity, uncharged
and with no image forces, $\epsilon_W=1$) is written as
\begin{subequations} \label{4.1}
\begin{eqnarray}
\Xi & = & \sum_{N=0}^{\infty} {z^{2N} \over (N!)^2} Q_{N,N}
\label{4.1a} \\
Q_{N,N} & = & \int_D \prod_{i=1}^N {\rm d}^2 p_i {\rm d}^2 n_i
~ W_{\beta}^{(N)}({\bf p},{\bf n}) \label{4.1b}
\end{eqnarray}
\end{subequations}
where $Q_{N,N}$ is the configuration integral and
\begin{equation} \label{4.2}
W_{\beta}^{(N)}({\bf p},{\bf n}) =
{ \prod_{(i<j)=1}^N \vert {\bf p}_i - {\bf p}_j \vert^{\beta} 
\vert {\bf n}_i - {\bf n}_j \vert^{\beta} \over
\prod_{i,j=1}^N \vert {\bf p}_i - {\bf n}_j \vert^{\beta} }
\end{equation}
denotes the interaction Boltzmann weight of $N$ particles of
charge $+1$ with coordinates $\{ {\bf p}_i \}_{i=1}^N$ and
$N$ particles of charge $-1$ with coordinates $\{ {\bf n}_i \}_{i=1}^N$.
From the explicit representations
\begin{subequations} \label{4.3}
\begin{eqnarray}
\langle {\hat n}({\bf r}) \rangle & = & {1\over \Xi}
\sum_{N=0}^{\infty} {z^{2N} \over (N!)^2}
\int_D \prod_{i=1}^N {\rm d}^2 p_i {\rm d}^2 n_i
~ W_{\beta}^{(N)}({\bf p},{\bf n}) \nonumber \\
& & \times \sum_{i=1}^N \left[ \delta({\bf r}-{\bf p}_i)
+ \delta({\bf r} - {\bf n}_i) \right] \label{4.3a} \\
\langle {\hat n}({\bf r}) {\hat n}({\bf r}') \rangle & = & 
{1\over \Xi} \sum_{N=0}^{\infty} {z^{2N} \over (N!)^2}
\int_D \prod_{i=1}^N {\rm d}^2 p_i {\rm d}^2 n_i
~ W_{\beta}^{(N)}({\bf p},{\bf n}) \nonumber \\
& & \times \sum_{i=1}^N \left[ \delta({\bf r}-{\bf p}_i)
+ \delta({\bf r} - {\bf n}_i) \right]
\sum_{j=1}^N \left[ \delta({\bf r}'-{\bf p}_i)
+ \delta({\bf r}' - {\bf n}_i) \right] \label{4.3b}
\end{eqnarray}
\end{subequations}
etc., one readily gets the important relations
\begin{subequations} \label{4.4}
\begin{eqnarray}
\int_D {\rm d}^2 r ~ n({\bf r}) & = & z {\partial \over \partial z} 
\ln \Xi \label{4.4a} \\
\int_D {\rm d}^2 r' ~ S_ n({\bf r},{\bf r}') & = & 
z {\partial \over \partial z} 
n({\bf r}) \label{4.4b}
\end{eqnarray}
\end{subequations}
etc.

The domain of interest in this section is the disk of radius
$R$, $D = \{ \vert {\bf r} \vert \le R \}$.
When one rescales the particle coordinates as
${\bf p}_i = R {\bf p}_i'$ and ${\bf n}_i = R {\bf n}_i'$,
it becomes evident that $\Xi(z,R)$ depends only on the
dimensionless combination $z^2 R^{4-\beta}$.
As a consequence,
\begin{equation} \label{4.5}
R {\partial \Xi \over \partial R} = \left( 2 - {\beta \over 2}
\right) z {\partial \Xi \over \partial z}
\end{equation}
If one accepts the expected value of $c=-1$ in the universal
$\ln R$-term of the large-$R$ expansion (\ref{1.24}) 
($\chi=1$ for the disk), $\ln \Xi = - \beta \Omega$ takes the form
\begin{equation} \label{4.6}
\ln \Xi(z,\beta,R) = (\pi R^2) \beta p - (2\pi R) \beta \gamma
- {1\over 6} \ln \left( z^{1/(2-\beta/2)} R \right)
+ {\rm const} + \cdots
\end{equation}
where
\begin{equation} \label{4.7}
\beta p = f_V(\beta) z^{1/(1-\beta/4)}
\end{equation}
$p$ is the bulk pressure, and
\begin{equation} \label{4.8}
\beta \gamma = f_S(\beta) z^{1/(2-\beta/2)}
\end{equation}
$\gamma$ is the surface tension.
The bulk particle density $n$ is given by
the equation
\begin{equation} \label{4.9}
n = z {\partial (\beta p) \over \partial z} 
= {1\over 1-(\beta/4)} \beta p
\end{equation}
which gives the equation of state (\ref{1.11}).

The particle density $n(r,R)$ depends on both the radius $R$ and
the distance $0\le r \le R$ from the centre of the disk.
By using the above mentioned scaling transformation of 
particle coordinates, one has
\begin{equation} \label{4.10}
n(r,R) = z^{1/(1-\beta/4)} g\left( z^{1/(2-\beta/2)} r,
z^{1/(2-\beta/2)} R \right)
\end{equation}
with an unknown function $g$.
The origin can be moved to the boundary via the
coordinate transformation $x = R-r$.
In order to distinguish between the two different functions,
we will use the obvious notation
\begin{equation} \label{4.11}
n_R(x) \equiv n(R-x,R)
\end{equation}
Similarly as in (\ref{4.10}),
\begin{equation} \label{4.12}
n_R(x) = z^{1/(1-\beta/4)} {\bar g}\left( z^{1/(2-\beta/2)} x,
z^{1/(2-\beta/2)} R \right)
\end{equation}
with an unknown function ${\bar g}$ different from $g$.
The definition (\ref{4.11}) implies
\begin{subequations} \label{4.13}
\begin{eqnarray}
{\partial n(r,R) \over \partial r} & = & -
{\partial n_R(x) \over \partial x} \label{4.13a} \\
{\partial n(r,R) \over \partial R} & = & 
{\partial n_R(x) \over \partial x} +
{\partial n_R(x) \over \partial R} \label{4.13b}
\end{eqnarray}
\end{subequations}
The transition from the disk of radius $R$ to a rectilinear
hard wall can be understood as the limiting $R\to\infty$
procedure, with $\lim_{R\to\infty} n_R(x) = n(x)$ being
the particle density at a finite distance $x\ge 0$ from
the plain hard wall.

It is evident that the derivative of
$\int_0^R \prod_{i=1}^N {\rm d}^2 p_i {\rm d}^2 n_i 
W_{\beta}^{(N)}({\bf p},{\bf n})$
with respect to $R$ is
$2\pi R \int_0^R \prod_{i=1}^N {\rm d}^2 p_i {\rm d}^2 n_i 
W_{\beta}^{(N)}({\bf p},{\bf n}) \sum_{i=1}^N \left[
\delta({\bf p}_i-{\bf R}) + \delta({\bf n}_i-{\bf R}) \right]$
where ${\bf R}$ is a position vector of any point at
the disk boundary.
Hereinafter, the integration range from 0 to $R$ formally means 
that $0\le \{ \vert {\bf p}_i\vert, \vert {\bf n}_i\vert \} \le R$.
As a result,
\begin{equation} \label{4.14}
{\partial \over \partial R} \ln \Xi = 2 \pi R n_R(0)
\end{equation}
With regard to eqs. (\ref{4.6}) and (\ref{4.9}), the finite-size
correction of the particle density at the boundary reads
\begin{equation} \label{4.15}
n_R(0) = \left( 1 - {\beta\over 4} \right) n -
{1\over R} (\beta \gamma) - {1\over 12\pi R^2} + \cdots
\end{equation}
The well-known contact theorem $n(0) = [1-(\beta/4)] n$
\cite{Henderson1,Henderson2} results as the $R\to\infty$ 
limit of eq. (\ref{4.15}).

By using the same procedure for $\Xi ~ n(r,R)$ one arrives at
\begin{equation} \label{4.16}
{\partial \over \partial R} \left[ \Xi ~ n(r,R) \right] =
\Xi R \int_{-\pi}^{\pi} {\rm d}\varphi \left[
\langle {\hat n}({\bf r}) {\hat n}(R,\varphi) \rangle
- n(r) \delta({\bf r}-{\bf R}) \right]
\end{equation}
where, in polar coordinates, ${\bf R}=(R,\varphi)$.
Using eq. (\ref{4.13b}) and after some simple algebra,
one can pass from (\ref{4.16}) to
\begin{equation} \label{4.17}
{\partial n_R(x) \over \partial x} +
{\partial n_R(x) \over \partial R} = R \int_{-\pi}^{\pi}
{\rm d}\varphi ~ U_n({\bf r},{\bf R}) 
\end{equation}
In the limit $R\to\infty$ and for a fixed $x$,
$n_R(x) = n(x) + O(1/R)$, so 
$\lim_{R\to\infty} \partial n_R(x)/\partial R \to 0$.
When one introduces the $y$-coordinate as follows $y=R \varphi$,
eq. (\ref{4.17}) takes the form
\begin{equation} \label{4.18}
{\partial n(x) \over \partial x} = \int_{-\infty}^{\infty}
{\rm d}y ~ U_n(0,x;y)
\end{equation} 
and we recover the two-dimensional version of the WLMB 
equation \cite{Lovett,Wertheim}.

We are now ready to take advantage of the exact grand-canonical
relation (\ref{4.4b}).
With regard to the scaling form of $n_R(x)$, eq. (\ref{4.12}),
it holds
\begin{equation} \label{4.19}
\left( 2 - {\beta\over 2} \right) z {\partial n_R(x) \over
\partial z} = 2 n_R(x) + x {\partial n_R(x) \over \partial x}
+ R {\partial n_R(x) \over \partial R}
\end{equation}
In the limit $R\to\infty$, the last term on the rhs of eq.
(\ref{4.19}) vanishes.
Inserting the resulting $z \partial n(x) / \partial z$ into
(\ref{4.4b}), one finally obtains
\begin{equation} \label{4.20}
\int_0^{\infty} {\rm d}x' \int_{-\infty}^{\infty} {\rm d}y ~
S_n(x,x';y) = {n(x) \over 1 - (\beta/4)} + {x\over 2[1-(\beta/4)]}
{\partial n(x) \over \partial x}
\end{equation}
This is the local form of the compressibility sum rule
in the presence of a plain hard wall.
Indeed, in the limit $x\to\infty$,
$n(x)\to n$ and $\partial n(x)/\partial x \to 0$ faster than
any power-law due to screening.
Consequently, in the bulk, (\ref{4.20}) reproduces
(\ref{1.22}).
It is straightforward to prove that eq. (\ref{4.20}) is
valid for any value of $\epsilon_W$.

\renewcommand{\theequation}{5.\arabic{equation}}
\setcounter{equation}{0}

\section{M\"obius invariance and sum rules}
We now consider the TCP confined to a two-dimensional
domain $D$, and study the action of the M\"obius conformal
transformation
\begin{equation} \label{5.1}
z' = {a z + b \over c z + d} , \quad \quad
z = {d z' - b \over - c z' + a}
\end{equation}
(with free complex parameters $ad - bc \ne 0$) of complex
particle coordinates $(z,{\bar z})$ on the configuration
integral $Q_{N,N}$ (\ref{4.1b}).
Under the conformal transformation (\ref{5.1}),
the domain $D$ is mapped onto the one denoted as $D'$,
the surface element ${\rm d}z {\rm d}{\bar z}$ is written as
\begin{equation} \label{5.2}
{\rm d}z {\rm d}{\bar z} = {(ad-bc)({\bar a}{\bar d}-
{\bar b}{\bar c}) \over (a-c z')^2 ({\bar a}-{\bar c}{\bar z}')^2}
{\rm d}z' {\rm d}{\bar z}'
\end{equation}
the square of the distance between two particles takes the form
\begin{equation} \label{5.3}
\vert z_i - z_j \vert^2 = {(ad-bc)({\bar a}{\bar d}-
{\bar b}{\bar c}) \over (a-c z_i') ({\bar a}-{\bar c}{\bar z}_i')
(a-c z_j') ({\bar a}-{\bar c}{\bar z}_j')} 
\vert z'_i - z'_j \vert^2
\end{equation}
so that the Boltzmann factor (\ref{4.2}) reads
\begin{eqnarray} \label{5.4}
W_{\beta}^{(N)}({\bf p},{\bf n}) & = & \left[ 
(ad-bc) ({\bar a}{\bar d}-{\bar b}{\bar c}) 
\right]^{-N\beta/2} \nonumber \\
& & \times \prod_{i=1}^N 
\left[ (a-c p_i') ({\bar a}-{\bar c}{\bar p}_i') \right]^{\beta/2}
\left[ (a-c n_i') ({\bar a}-{\bar c}{\bar n}_i') \right]^{\beta/2}
~ W_{\beta}^{(N)}({\bf p}',{\bf n}')
\end{eqnarray}  
Finally, putting $c={\bar c}=1$ and $a=r_0$, ${\bar a}={\bar r}_0$,
one finds
\begin{eqnarray} \label{5.5}
\int_D \prod_{i=1}^N {\rm d}^2 p_i {\rm d}^2 n_i
~ W_{\beta}^{(N)}({\bf p},{\bf n}) & = & \left[
(b-r_0 d) ({\bar b}-{\bar r}_0 {\bar d})
\right]^{N[2-(\beta/2)]} \nonumber \\
& & \times \int_{D'} \prod_{i=1}^N 
{{\rm d}^2 p_i \over \vert {\bf r}_0 - {\bf p}_i \vert^{4-\beta}} 
{{\rm d}^2 n_i \over \vert {\bf r}_0 - {\bf n}_i \vert^{4-\beta}}
~ W_{\beta}^{(N)}({\bf p},{\bf n})
\end{eqnarray}
This means that the configuration integral of the TCP Coulomb
system confined to the domain $D$ is mappable onto the one
of the TCP Coulomb system (with the same number of charged particles
and at the same inverse temperature $\beta$) confined to the new
domain $D'$ and being under an additional action of an external 
gravitational source with specific coupling strength equal to $4-\beta$.
Since in the stability range of $\beta<2$ the gravitational
coupling $4-\beta>2$ and the lhs of eq. (\ref{5.5}) is finite,
the formalism must ensure that, for any values of free
complex parameters $b$ and $d$ such that $(b/d)\ne r_0$, the
gravitational source is localized outside of the domain $D'$
in order to prevent the divergence of the rhs of eq. (\ref{5.5}) 
due to the gravitational collapse.  
According to the definition (\ref{4.1a}), the relation between
the configuration integrals (\ref{5.5}) also implies an analogous
relation between the grand partition functions,
\begin{subequations} \label{5.6}
\begin{eqnarray}
\Xi(z,\beta,D) & = & \Xi({\tilde z},\beta,D'\vert {\bf r}_0)
\label{5.6a} \\
{\tilde z} & = & z \left[ (b-r_0 d)({\bar b}-{\bar r}_0{\bar d})
\right]^{1-(\beta/4)} \label{5.6b}
\end{eqnarray}
\end{subequations} 

Taking for the domain a disk with its center at the origin,
$D = \{ \vert r \vert \le R \}$,
let us set $b=R^2$ and $d={\bar r}_0$ $(r_0 {\bar r}_0 \ne R^2)$,
besides the already chosen $c=1$ and $a=r_0$, so that the M\"obius
conformal transformation (\ref{5.1}) takes the form
\begin{equation} \label{5.7}
z' = {r_0 z + R^2 \over z + {\bar r}_0}, \quad \quad
z = {{\bar r}_0 z' - R^2 \over -z' + r_0}
\end{equation}
It is straightforward to show that under this transformation
the disk $z{\bar z}\le R^2$ maps onto a ``dual'' domain $D'$
defined by the inequality
\begin{equation} \label{5.8}
(R^2 - r_0 {\bar r}_0) (R^2 - z' {\bar z}') \le 0
\end{equation}
If the gravitational point is outside of the disk, i.e.
$r_0 {\bar r}_0 > R^2$, the transformation (\ref{5.7})
maps the disk onto itself.
Relation (\ref{5.5}) results into
\begin{eqnarray} \label{5.9}
\int_0^R \prod_{i=1}^N {\rm d}^2 p_i {\rm d}^2 n_i
~ W_{\beta}^{(N)}({\bf p},{\bf n}) & = & 
\left( r_0{\bar r}_0 - R^2 \right)^{N(4-\beta)} \nonumber \\
& & \times \int_0^R \prod_{i=1}^N 
{{\rm d}^2 p_i \over \vert {\bf r}_0 - {\bf p}_i \vert^{4-\beta}} 
{{\rm d}^2 n_i \over \vert {\bf r}_0 - {\bf n}_i \vert^{4-\beta}}
~ W_{\beta}^{(N)}({\bf p},{\bf n})
\end{eqnarray}
If $r_0{\bar r}_0 < R^2$, the disk is mapped onto its complement
in the two-dimensional space and eq. (\ref{5.5}) gives
\begin{eqnarray} \label{5.10}
\int_0^R \prod_{i=1}^N {\rm d}^2 p_i {\rm d}^2 n_i
~ W_{\beta}^{(N)}({\bf p},{\bf n}) & = & 
\left( R^2 - r_0{\bar r}_0 \right)^{N(4-\beta)} \nonumber \\
& & \times \int_R^{\infty} \prod_{i=1}^N 
{{\rm d}^2 p_i \over \vert {\bf r}_0 - {\bf p}_i \vert^{4-\beta}} 
{{\rm d}^2 n_i \over \vert {\bf r}_0 - {\bf n}_i \vert^{4-\beta}}
~ W_{\beta}^{(N)}({\bf p},{\bf n})
\end{eqnarray}
In both cases, the gravitational source lies outside of the
dual domain, as is required by the condition of stability (see
the comment in the above paragraph).
There exists a special M\"obius transformation which maps the disk
directly onto a half-plane, but we will not discuss this case.

Although the present paragraph will not be used in the following,
let us remark that the self-duality of the disk system (\ref{5.9})
produces a relation between the grand partition functions
of type (\ref{5.6}), with $D'=D={\rm disk}$ and
${\tilde z} = z(r_0 {\bar r}_0 - R^2)^{2-(\beta/2)}$.
Since $\Xi(z,\beta,R)$ depends on $\beta$ and the combination
$z^2 R^{4-\beta}$, $\Xi({\tilde z},\beta,R\vert {\bf r}_0)$ 
will depend on $\beta$ and the combination 
${\tilde z}^2 R^{4-\beta} / (r_0 {\bar r}_0 - R^2)^{4-\beta}$.
Let $x = \vert r_0 \vert - R$ be the distance of the gravitational
point from the surface of the disk.
The consequent ratio $R/(2xR+x^2)$ diverges in the limit
$R\to\infty$, $x\to 0$.
In this case, the asymptotic formula (\ref{4.6}) can be applied.
When $x$ is finite, the asymptotic formula (\ref{4.6}) is applicable 
only when one multiplies $z$ by an appropriate $R$-dependent
constant via changing the zero reference energy of the gravitational
potential.

We will now use the self-dual relation (\ref{5.9}) to derive
a new sum rule (\ref{5.14}) and its more detailed version (\ref{5.17})
for the density structure function $S_n$ in
the case of the disk geometry.
First, we rewrite the rhs of (\ref{5.9}) as follows
\begin{equation} \label{5.11}
\int_0^R \prod_{i=1}^N {\rm d}^2 p_i \left[ 
{1-R^2/(r_0{\bar r}_0) \over (1-p_i/r_0) (1-{\bar p}_i/{\bar r}_0)}
\right]^{2-(\beta/2)}
{\rm d}^2 n_i \left[ 
{1-R^2/(r_0{\bar r}_0) \over (1-n_i/r_0) (1-{\bar n}_i/{\bar r}_0)}
\right]^{2-(\beta/2)} W_{\beta}^{(N)}({\bf p},{\bf n})
\end{equation}
Then, since $R^2/(r_0{\bar r}_0)<1$, $\vert p_i/r_0 \vert <1$ and
$\vert n_i/r_0 \vert <1$, we perform the large-$\vert r_0\vert$
expansion
\begin{eqnarray} \label{5.12}
\left[ {1-R^2/(r_0{\bar r}_0) \over (1-z/r_0) (1-{\bar z}/{\bar r}_0)}
\right]^{2-(\beta/2)} & = & 1 + 
\left( 2 - {\beta\over 2} \right){z\over r_0} + 
\left( 2 - {\beta\over 2} \right){{\bar z}\over {\bar r}_0} 
\nonumber \\ & & -
\left( 2 - {\beta\over 2} \right){R^2 \over r_0 {\bar r}_0} +
\left( 2 - {\beta\over 2} \right)^2 {z {\bar z} \over r_0 {\bar r}_0}
+ \cdots 
\end{eqnarray}
for each $z = \{ p_i, n_i \}$.
The term of order $(r_0{\bar r}_0)^0$ in (\ref{5.11}) exactly
reproduces the lhs of (\ref{5.9}), and the coefficients of 
higher-order terms in inverse powers of $r_0$ and ${\bar r}_0$ 
must be identically equal to zero.
The terms $1/r_0$ and $1/{\bar r}_0$ trivially vanish.
The $1/(r_0{\bar r}_0)$ term vanishes when
\begin{equation} \label{5.13}
\int_0^R \prod_{i=1}^N {\rm d}^2 p_i {\rm d}^2 n_i
\sum_{j,k=1}^N (p_j+n_j) ({\bar p}_k+{\bar n}_k)
W_{\beta}^{(N)}({\bf p},{\bf n}) =
{2 N R^2 \over 2 - (\beta/2)} \int_0^R
\prod_{i=1}^N {\rm d}^2 p_i {\rm d}^2 n_i
W_{\beta}^{(N)}({\bf p},{\bf n})
\end{equation} 
Within the grand-canonical formalism (\ref{4.1})-(\ref{4.3}),
the equality (\ref{5.13}) is equivalent to
\begin{equation} \label{5.14}
\int_0^R {\rm d}^2r \int_0^R {\rm d}^2 r' ~ 
{\bf r}\cdot {\bf r}' S_n^{(R)}({\bf r},{\bf r}') 
= {R^2 \over 2- (\beta/2)}
z {\partial \over \partial z} \ln \Xi(z,\beta,R)
\end{equation}

Everything that has been done in the case of the configuration
integral can be adapted to the particle density.
Let us ``rotate'' the M\"obius transformation (\ref{5.7})
around the origin by the multiplication factor $({\bar r}_0/r_0)$
in order to obtain the identity $z'=z$ in the limit
$\vert r_0\vert \to \infty$, and denote by $a$ the ratio
$R/r_0$ whose absolute value is smaller than 1,
\begin{equation} \label{5.15}
z' = {z + a R \over 1 + ({\bar a}z/R)}
\end{equation}
For a given point ${\bf r} = (r,{\bar r})$ in the interior
of the disk, the previously developed formalism implies
\begin{eqnarray} \label{5.16}
& & \int_0^R \prod_{i=1}^N {\rm d}^2 p_i {\rm d}^2 n_i
\sum_{j=1}^N \left[ \delta({\bf r}-{\bf p}_j) +
\delta({\bf r}-{\bf n}_j) \right] 
W_{\beta}^{(N)}({\bf p},{\bf n}) \nonumber \\
& & = {(1-a{\bar a})^2 \over (1+{\bar a}r/R)^2 (1+a{\bar r}/R)^2}
\int_0^R \prod_{i=1}^N {\rm d}^2 p_i
\left[ {1- a{\bar a} \over (1- {\bar a}p_i/R) (1- a {\bar p}_i/R)}
\right]^{2-(\beta/2)} {\rm d}^2 n_i \nonumber \\
& & \times \left[ {1- a{\bar a} \over (1- {\bar a}n_i/R) 
(1- a {\bar n}_i/R)} \right]^{2-(\beta/2)}
\sum_{j=1}^N \left[ \delta({\bf r}'-{\bf p}_j) +
\delta({\bf r}'-{\bf n}_j) \right] 
W_{\beta}^{(N)}({\bf p},{\bf n})
\end{eqnarray}
When one expands the rhs of (\ref{5.16}) in powers of
$a = a_x + {\rm i}a_y$, $\vert a\vert <1$, the requirement
of the nullity of the coefficients attached to $a_x$ and $a_y$
leads to the relation
\begin{equation} \label{5.17}
(4-\beta) \int_0^R {\rm d}^2 r' ~ {\bf r} \cdot {\bf r}'
S_n^{(R)}({\bf r},{\bf r}') = 4 r^2 n(r,R) +
(r^2-R^2) r {\partial \over \partial r} n(r,R)
\end{equation}
The new sum rule (\ref{5.14}) results from the more detailed
eq. (\ref{5.17}) when one integrates $\int_0^R {\rm d}^2 r$ 
both sides of the latter, then uses an integration per partes
and finally applies the equality (\ref{4.4a}).

This new sum rule (\ref{5.17}) has a consequence for the density
near the boundary.
Let us divide eq. (\ref{5.17}) by $\vert {\bf r} \vert$,
and move the origin to the boundary via the coordinate
transformation $r=R-x$ described in section 4.
Then, using eq. (\ref{4.19}), and in the $R\to\infty$
limit, we find after some simple algebra that
\begin{equation} \label{5.18}
n_R(x) =  n(x) - {1\over 2R} \left\{ (4-\beta)
\int_0^{\infty} {\rm d}x' ~ x' \int_{-\infty}^{\infty}
{\rm d}y ~ S_n(x,x';y) - 4 x n(x) - 
x^2 {\partial n(x) \over \partial x} \right\} + \cdots 
\end{equation}
This formula determines the leading correction of the density
$n_R(x)$, at a fixed distance $x$ from the wall, with respect 
to its asymptotic $R\to\infty$ value $n(x)$ due to the curvature
of the confining disk domain.
The quantities on the rhs of (\ref{5.18}) are the ones
evaluated for the rectilinear hard wall.

The asymptotic formula (\ref{5.18}) can be explicitly
checked at the boundary, where it gives
\begin{equation} \label{5.19}
n_R(0) = n(0) - {4-\beta \over 2 R} 
\int_0^{\infty} {\rm d}x ~ x \int_{-\infty}^{\infty} {\rm d}y
~ S_n(0,x;y) + \cdots
\end{equation}
Multiplying eq. (\ref{4.18}) by $x$, integrating then over
$x$ from 0 to $\infty$ and performing an integration per partes,
one finds
\begin{equation} \label{5.20}
- \int_0^{\infty} {\rm d}x \left[ n(x)-n \right] =
\int_0^{\infty} {\rm d}x ~ x \int_{-\infty}^{\infty} {\rm d}y
\left[ S_n(0,x;y) - n(x) \delta(x) \delta(y) \right]
\end{equation}
On the other hand, according to the relation (\ref{4.4a}),
it holds for the disk
\begin{equation} \label{5.21}
\int_0^R {\rm d}^2 r \left[ n(r,R) - n \right] =
z {\partial \ln \Xi \over \partial z} - n \pi R^2
\end{equation}
Assuming the large-$R$ expansion (\ref{4.6}) and
passing to the boundary-distance variable $x=R-r$,
eq. (\ref{5.21}) takes the form
\begin{equation} \label{5.22}
2\pi \int_0^R {\rm d}x (R-x) \left[ n_R(x)-n \right]
= - (2\pi R) z {\partial (\beta \gamma) \over \partial z}
- {1\over 12 (1-(\beta/4))} + \cdots
\end{equation} 
In the limit $R\to \infty$, this equation implies
the obvious boundary relation
\begin{equation} \label{5.23}
\int_0^{\infty} {\rm d}x \left[ n(x)-n \right] =
- z {\partial (\beta \gamma)\over \partial z}
\end{equation}
Taking into account the $z$-dependence of $\beta\gamma$
(\ref{4.8}) in (\ref{5.23}) and considering the previously
derived relations (\ref{5.19}) and (\ref{5.20}), 
large-$R$ asymptotics (\ref{4.15}) is reproduced correctly up to 
the $(1/R)$-term, which confirms the validity of (\ref{5.19}).

The asymptotic formula (\ref{5.18}) implies a generalization
of the second moment density sum rule (\ref{1.23}) to the
case of a rectilinear wall.
Indeed, the local version of the compressibility sum rule 
(\ref{4.20}) can be used for rewriting (\ref{5.18}) as
\begin{equation} \label{5.24}
n_R(x) = n(x) - {1\over 2R} \left\{ (4-\beta) \int_0^{\infty}
{\rm d}x' (x'-x) \int_{-\infty}^{\infty} {\rm d}y ~
S_n(x,x';y) + x^2 {\partial n(x) \over \partial x} \right\}
+ \cdots
\end{equation}
We will suppose that the half-infinite Coulomb system has good 
screening properties into the bulk, i.e., $[n(x)-n]$
decays faster than any inverse-power law as $x\to\infty$
and all moments $\int_0^{\infty}{\rm d}x ~ x^i [n(x)-n]$ exist.
From eqs. (\ref{5.22}) and (\ref{5.23}), one then gets in
the limit of large $R$
\begin{equation} \label{5.25}
2\pi R \int_0^{\infty} {\rm d}x \left[ n_R(x) - n(x) \right]
- 2 \pi \int_0^{\infty} {\rm d}x ~ x \left[ n(x)-n \right]
= - {1\over 12 (1-(\beta/4))}
\end{equation}
Inserting (\ref{5.24}) into (\ref{5.25}) and after some algebra
we arrive at
\begin{equation} \label{5.26}
\int_0^{\infty} {\rm d}x \int_0^{\infty} {\rm d}x' (x'-x)
\int_{-\infty}^{\infty} {\rm d}y ~ S_n(x,x';y) =
{1\over 48 \pi (1-(\beta/4))^2}
\end{equation}

Actually, it can be shown that the boundary sum rule (\ref{5.26}) 
can be derived directly from the second-moment sum rule 
in the bulk (\ref{1.23}) by using a simple assumption. 
Indeed, $S_n(x,x';y)$ can be decomposed as the sum of the
bulk structure factor plus a surface term:
\begin{equation} \label{5.27}
S_n(x,x';y)=S_n^{{\rm bulk}}(r)+S_n^{{\rm surface}}(x,x';y)
\end{equation}
where $r=[(x-x')^2+y^2]^{1/2}$ and $S_n^{{\rm bulk}}(r)$ is the bulk 
density structure factor appearing in (\ref{1.23}) 
[(\ref{5.27}) is just a definition of $S_n^{{\rm surface}}$]. 
When this decomposition is used in the lhs of 
(\ref{5.26}), assuming that $S_n^{{\rm surface}}(x,x';y)$ has a fast
decay when $x$ or $x'$ or both go to infinity (see for instance the
explicit expressions in the weak-coupling or $\beta=2$ cases considered
in sections 3.1 and 3.2), the corresponding integral is
absolutely convergent, the order of the integrations on $x$ and $x'$ can
be freely interchanged, and since $(x-x')S_n^{{\rm surface}}(x,x';y)$ is 
odd under the interchange of $x$ and $x'$, this integral vanishes.
This reasoning however does not apply to the contribution of the 
bulk structure factor which does not decay to 0 when both $x$ and $x'$ 
go to infinity for a fixed value of $x-x'$; 
the corresponding integral is not absolutely convergent, 
and the order of the integrations cannot be changed. 
Thus, only the bulk part of $S_n$ contributes to the lhs $I$
of (\ref{5.26}) which can be rewritten as
\begin{equation} \label{5.28}   
I=\int_0^{\infty}{\rm d}x\int_{-x}^{\infty}{\rm d}s ~ s
\int_{-\infty}^{\infty}{\rm d}y ~ S_n^{{\rm bulk}}(|s|;y)
\end{equation}
where we have used the integration variable $s=x'-x$ rather than $x'$
and made explicit that $S_n^{{\rm bulk}}$ depends on $x$ and $x'$ only
through $|s|$. 
Performing the integration on $x$ per partes gives
\begin{equation} \label{5.29}
I = x \int_{-x}^{\infty}{\rm d}s ~ s \int_{-\infty}^{\infty}
{\rm d}y ~ S_n^{{\rm bulk}}(|s|;y) \big\vert_{x=0}^{x=\infty} +
\int_0^{\infty}{\rm d}x\int_{-\infty}^{\infty}{\rm d}y\,x^2
S_n^{{\rm bulk}}(|x|;y)
\end{equation}
Because $S_n^{{\rm bulk}}$ has a fast (exponential) decay at
infinity, the first term in the rhs of (\ref{5.29}) vanishes.
Since $S_n^{{\rm bulk}}(|x|;y)$ is a function of $r=(x^2+y^2)^{1/2}$ 
only, $I$ is 1/4 of the lhs of (\ref{1.23}), and (\ref{1.23}) results 
into (\ref{5.26}).
It should be noted that the above derivation of the boundary sum rule
(\ref{5.26}) still holds for an arbitrary value of the wall dielectric
constant $\epsilon_W$, including the special cases $\epsilon_W=0$ and 
$\epsilon_W=\infty$. 

\renewcommand{\theequation}{6.\arabic{equation}}
\setcounter{equation}{0}

\section{Universal finite-size correction for the disk}
In most considerations of the previous two sections,
we have assumed the validity of the large-$R$ expansion
of $\ln \Xi$ (\ref{4.6}) for the disk, in particular
the value $-1/6$ of the coefficient of the universal
$\ln R$-term.
Here, we prove this assumption.

Using the formula $\vert {\bf r}-{\bf r}'\vert^2
= \vert {\bf r}\vert^2 + \vert {\bf r}'\vert^2
-2 {\bf r}\cdot {\bf r}'$, we write
\begin{equation} \label{6.1}
\int_0^R {\rm d}^2 r \int_0^R {\rm d}^2 r' 
\vert {\bf r}-{\bf r}'\vert^2 S_n^{(R)}({\bf r},{\bf r}') 
= 2 z {\partial\over \partial z}
\left[ \int_0^R {\rm d}^2 r \vert {\bf r} \vert^2 n(r,R)
- {R^2 \over 2-(\beta/2)} \ln \Xi \right] 
\end{equation}
where we have used the relation (\ref{4.4b}) and 
the new sum rule (\ref{5.14}).
Let us consider the large-$R$ expansion (\ref{4.6}) in a
general form
\begin{equation} \label{6.2}
\ln \Xi = (\pi R^2) \left( 1-{\beta\over 4} \right) n
- (2\pi R) \beta \gamma + f
\end{equation}
where $n\propto z^{1/(1-\beta/4)}$, $\beta \gamma \propto
z^{1/(2-\beta/2)}$ and $f$ is an as-yet-undetermined function
of $\beta$ and the combination $z^2 R^{4-\beta}$. 
The representation (\ref{6.2}) when combined with (\ref{5.21})
implies
\begin{equation} \label{6.3}
\beta \gamma = {1-(\beta/4) \over \pi R} \left\{
z {\partial f \over \partial z} - \int_0^R {\rm d}^2 r
\left[ n(r,R)-n \right] \right\}
\end{equation}
Substituting $\beta\gamma$ in (\ref{6.2}) by the expression
(\ref{6.3}) and then inserting $\ln \Xi$ into (\ref{6.1}),
one obtains
\begin{eqnarray} \label{6.4}
{1\over \pi R^2} \int_0^R {\rm d}^2 r \int_0^R {\rm d}^2 r'
\vert {\bf r} - {\bf r}' \vert^2 S_n^{(R)}({\bf r},{\bf r}')
& = & - {2\over \pi} z {\partial \over \partial z} 
\int_0^R {\rm d}^2 r \left( 1 - {r^2\over R^2} \right)
\left[ n(r,R) - n \right] \nonumber \\
& & - {1\over \pi (1-(\beta/4))} z {\partial \over \partial z} f
+ {2\over \pi} \left( z {\partial \over \partial z} \right)^2 f
\end{eqnarray}
This equation, which is exact for an arbitrary disk radius $R$,
will now be analyzed in the limit $R\to\infty$.
In that limit, since $S_n^{(R)}({\bf r},{\bf r}')$ decays
along the boundary like $1/\vert {\bf r}-{\bf r}' \vert^4$
(see section 3), there is no relevant surface contribution 
to the integral on the lhs of (\ref{6.4}), and we can make use of 
the bulk sum rule (\ref{1.23}) derived in \cite{Jancovici5},
\begin{equation} \label{6.5}
\lim_{R\to\infty} {1\over \pi R^2} \int_0^R {\rm d}^2 r 
\int_0^R {\rm d}^2 r' \vert {\bf r} - {\bf r}' \vert^2 
S_n^{(R)}({\bf r},{\bf r}') = 
{1\over 12 \pi (1-(\beta/4))^2}
\end{equation}
As concerns the rhs of (\ref{6.4}), the unknown integral
over the disk area can be represented in the 
$x=R-r$ coordinate as
\begin{equation} \label{6.6}
\int_0^R {\rm d}^2 r \left( 1 - {r^2\over R^2} \right)
\left[ n(r,R) - n \right] = 4 \pi \int_0^R {\rm d}x ~
x \left( 1 - {x\over R} \right) \left( 1 + {x\over 2 R} \right)
\left[ n_R(x) - n \right]
\end{equation}
In the limit $R\to\infty$, due to good screening properties
of the plasma, only the term $\propto \int_0^{\infty} {\rm d}x ~ x
[n(x)-n]$ survives.
Based on the scaling analysis of section 3, since $n(x) = n
{\bar g}(\sqrt{n}x)$, this integral is dimensionless, and therefore
its derivative with respect to $z$ vanishes.
Consequently, without the integral on the rhs of eq. (\ref{6.4}),
the solution of that equation combined with (\ref{6.5}) reads
\begin{equation} \label{6.7}
f = - {1\over 6} \ln \left( R z^{1/(2-\beta/2)} \right)
+ {\rm const} + \cdots
\end{equation}
in full agreement with the expected universal large-$R$ 
behavior (\ref{4.6}).

\renewcommand{\theequation}{7.\arabic{equation}}
\setcounter{equation}{0}

\section{Concluding remarks}
The pecularities of the charge correlation in a Coulomb fluid
are determined by the screening effect caused by the long-ranged
tail of the Coulomb potential.
The density correlation function in a charge-symmetrical two-component 
plasma, decoupled from the charge one at the level of the OZ equation 
[see relations (\ref{2.1})], does not feel this tail directly.
Yet, the density correlation bears some similarity with the
charge correlation.
Although the density correlation decays always faster than the 
charge one, it still exhibits an algebraically slow decay 
$\sim \vert {\bf y} \vert^{- 2\nu}$ (\ref{3.4}) at large distances 
$\vert {\bf y} \vert$ along a rectilinear hard wall 
in any dimension $\nu$.
An analogous phenomenon for the charge correlation, which
decays like $\sim \vert {\bf y} \vert^{-\nu}$ (\ref{3.3}),
is related to an asymmetry of the screening cloud near the wall.

In this paper, we have concentrated mainly on the two-dimensional TCP 
of pointlike charges.
The form-factor analysis of the equivalent sine-Gordon model
was done in the bulk case.
In contrast to the charge correlation, the density one exhibits
a discontinuity of the slope of the correlation length, namely
at point $\beta=1$ [see relations (\ref{2.31}) and (\ref{2.32})].
At the collapse point $\beta=2$, as seen in eqs. (\ref{2.10})
and (\ref{2.11}), the charge and density correlation lengths 
become the same and the difference in the asymptotic decay is 
only in the inverse-power law prefactors.

The TCP in contact with a rectilinear hard wall of dielectric
constant $\epsilon_W$ was mapped onto an integrable boundary
sine-Gordon model in two cases: $\epsilon_W \to \infty$
(ideal metal wall, Dirichlet boundary condition) \cite{Samaj2}
and $\epsilon_W=0$ (ideal dielectric wall, Neumann boundary
condition) \cite{Samaj3}.
Interestingly, just in these two cases the charge and density
correlation functions decay exponentially also along the boundary.
The next potentially integrable case corresponds to the plain
hard wall with $\epsilon_W=1$: at least, the explicit solution 
is available at the free-fermion point \cite{Cornu2}.
For this case, we have suggested an explicit form of the
asymptotic characteristics along the wall, $f_{\rho}(x,x')$
(\ref{3.15}) and $f_n(x,x')$ (\ref{3.16}), as a natural
interpolation between the exact results in the $\beta\to 0$
limit and at $\beta=2$.
The universal value of the prefactor $=1/(2\pi^2)$ in $f_n$
deserves attention.
To check the validity of our conjectures, one should go
beyond the Debye-H\"uckel limit.

The scaling analysis of section 4 leads to a new local
version of the compressibility sum rule in the presence
of a hard wall (\ref{4.20}).

In the crucial section 5, a symmetry of the TCP confined to
a disk with respect to the M\"obius conformal transformation
of particle coordinates, which induces a gravitational source
point interacting with the particles, is established.
This symmetry is of interest by itself: it produces a path
between the pure Coulomb system and the one under the action
of a gravitational point.
The gravitational point acts in the same way on positive
and negative charges, so that the total particle density is
relevant.
As a result, the new sum rule (\ref{5.14}) for the density
correlation function and its local version (\ref{5.17}) are
derived for the disk geometry of the Coulomb system.
The finite-size formula (\ref{5.18}) for the particle density
at a fixed distance from the wall and the boundary version
(\ref{5.26}) of the bulk second moment of $S_n$ (\ref{1.23})
also deserve attention. 

The new density sum rule (\ref{5.14}) is used in section 6 to derive 
the universal finite-size correction term of the grand potential
for the disk geometry.
We would like to stress a fundamental difference between the
derivation of the universal finite-size correction term
for a system which is finite in each direction (a disk or a 
sphere, a {\it density} sum rule is relevant) and for a system
which is infinite at least in one direction (a strip or a
$\nu$-dimensional slab, a {\it charge} sum rule is relevant
\cite{Jancovici7,Jancovici8}).

In most cases, we have restricted ourselves to the
symmetrical-charge TCP and $\epsilon_W=1$ in order to
maintain the clarity of presentation.
The extension of the formalism to the asymmetric TCP
and arbitrary $\epsilon_W$ is usually straightforward. 

\section*{Acknowledgments}
The stay of Ladislav \v Samaj in LPT Orsay is supported by a NATO
fellowship. A partial support by Grant VEGA 2/7174/20 is acknowledged.

\end{document}